\documentclass[twocolumn]{revtex4}

\usepackage{graphicx}
\usepackage{subfigure}
\usepackage{ulem}
\usepackage{sidecap}
\usepackage{color}
\usepackage{appendix}

\makeatletter

\graphicspath{{./figs2/}}

\begin{document}

\title{Effects of community structure on epidemic spread in an adaptive network}

\author{Ilker Tunc}

\affiliation{Department of Applied Science, College of William and Mary,
Williamsburg, VA 23187}

\author{Leah B.~Shaw}

\affiliation{Department of Applied Science, College of William and Mary,
Williamsburg, VA 23187}

\begin{abstract}
When an epidemic spreads in a population, individuals may adaptively change
the structure of their social
contact network to reduce risk of infection. Here we study the spread of
an epidemic on an adaptive network
with community structure. We model the effect of two communities with different average degrees.  The disease model is susceptible-infected-susceptible (SIS), and adaptation is rewiring of links between susceptibles and infectives.
The bifurcation structure is obtained, and a mean field model is developed that accurately predicts the steady state behavior of the system.  We show that an epidemic can alter the community structure.
\end{abstract}

\maketitle

\section{Introduction}

In recent years, networks have been widely used in modeling a variety of social, technological, and biological systems \cite{Albert2002,Dorogovtsev2001,Newman2003}.  One major application is modeling the spread of an epidemic on a social network (e.g., \cite{Pastor-Satorras2001a,Kuperman2001,May2001,Pastor-Satorras2001,Newman2002}). In these models, typically the network structure is assumed static, while the infection status of the nodes changes dynamically.

During an epidemic, people may tend to avoid social connections with infected individuals
\cite{Gross2006,Schwartz2010}. This system can be considered an adaptive network, where the
node dynamics affects the network topology, which then affects future changes
in node status \cite{Gross2008a}. Epidemic spreading in adaptive network models
with avoidance behavior has been studied previously
\cite{Gross2006,Zanette2007,Gross2008a,Shaw2008a,VanSegbroeck2010,Marceau2010}, with avoidance frequently implemented via susceptible nodes rewiring their links away from infected neighbors and towards other non-infected nodes. Changes
in bifurcation structure have been observed, including the existence of bistable regimes with endemic and disease free states both stable.

One of the most important features of a social network is community
structure \cite{Newman2004}. The strength of the community structure can be
quantified by using a modularity measure \cite{Newman2004}, which for a random
network will be close to zero, and  will be close to one for a strong
community structure.

Studies of epidemics with community structure have focused mainly on static network geometries, including scale-free
\cite{Yan2007,Huang2007,Chu2009}, small-world
\cite{Zhao2007} and random networks \cite{Liu2005a}.
It has been found that community structure can either decrease \cite{Huang2007} or increase \cite{Liu2005a} infection prevalence, depending on details of the model.
Further, epidemics can synchronize across communities if there are sufficient connections between communities \cite{Yan2007,Zhao2007}.
In a dynamic but not adaptive example, communities of mobile agents were studied \cite{Zhou2009,Xia2009}, and dynamic hopping of agents between communities was able to produce sustained infection in communities that were below the epidemic threshold if other communities were above threshold.

Adaptive networks with community structure have been studied only rarely.
In \cite{Sun2007}, the authors considered an adaptive scale-free
network with community structure in which neighbors of an infected node can
move to other communities with a certain  probability. Infection levels were
reduced compared to the case without adaptation, but the adaptation mechanism
did not preserve the community structure as measured by modularity. In \cite{Wang2011}, the authors introduced a model very similar in structure to the one we will consider here.  However, their focus was to study an adaptive epidemic system with two types of agents.  They varied the within-type and cross-type link rewiring rates and infection rates and determined their effects on the size of the bistability region.  For certain parameter choices the endemic steady state would have a community structure, but the resulting structure was not characterized in this study.

In this paper, we extend the adaptive susceptible-infected-susceptible (SIS) model of \cite{Gross2006} to a network with two communities.  In contrast to previous studies \cite{Sun2007,Wang2011}, we allow the communities to have different average degree.  We define rewiring rules such that the community structure is preserved if links between susceptibles and infectives are uniformly distributed.  We directly simulate the stochastic network system and derive a lower dimensional mean field, based on a moment closure approximation, that accurately predicts the bifurcation structure of the full system.  In Section \ref{sec:model}, we define the model and introduce the mean field equations.  Results in the absence of adaptation are presented in Subsection \ref{subsec:static}.  In Subsection \ref{subsec:adaptive}, we show the effects of adaptation on the bifurcation structure and on the network geometry.  Section \ref{sec:conclusions} concludes.


\section{Model}

\label{sec:model}

We study a susceptible-infective-susceptible (SIS) model on an adaptive
network having two communities. The communities are labeled A and B and consist of $N_a$ and $N_b$ nodes, respectively.  Here $N_a=N_b=5000$.
We use two probability parameters to generate an initial network with two
communities by creating links.  Parameter $d \in [0,1]$ determines the asymmetry in
average degree of the communities, and $f \in [0,1]$ determines the number of links
between communities.  Links are created as follows.  With probability $d$ we
choose a node among the $N_a$ nodes in community A (otherwise choosing a
node in community B), and with probability $f$ its neighbor is chosen at
random from the opposite community as the first node (otherwise choosing from the
same community). This process is repeated until a total of $K$ links are
created.  Self links and multiple links are disallowed. Then a fraction $(1-f)d$ of the links are AA, $(1-f)(1-d)$ are BB, and
 $f$ are AB.  The average degrees in
communities A and B are $\langle k_a \rangle=\left[2(1-f)d+f\right]K/N_a$ and  $\langle k_b \rangle=\left[2(1-f)(1-d)+f\right]K/N_b$,
respectively.  Thus the communities are symmetric when $d=0.5$.  We will focus
here on the case $d>0.5$, so community A will have higher connectivity than
B. The total number of cross links between communities is $fK$.

We define node dynamics as in \cite{Gross2006}. A susceptible (S) node becomes
infected with rate $pN_{inf}$, where $N_{inf}$ is the number of infected
neighbors the node has and $p$ is the infection rate. An infected (I) node
recovers with recovery rate $r$.  One of these rates can be eliminated by
rescaling time, so it is sufficient to treat $r$ as fixed.  We fix $r=0.002$
throughout the paper as in previous studies \cite{Gross2006,Shaw2008a} .

Network adaptation in the form of avoidance behavior is introduced by allowing
susceptible-infected links to rewire with rate $w$ to susceptible-susceptible
links, as in \cite{Gross2006}.  However, the rewiring must be
adjusted to retain the desired community structure.  This is done by choosing
the susceptible node's new neighbor from one or the other community with
appropriate probabilities.  An S node having an infected neighbor
rewires to an S node in the same community as itself with probability $\alpha$
if the S node is in community A and with probability $\beta$ if
the S node is in community B.  Otherwise a neighbor in the other
community is selected. In order to retain the community structure, we set
$\alpha=\frac{2(1-f)d}{2(1-f)d+f}$ and
$\beta=\frac{2(1-f)(1-d)}{2(1-f)(1-d)+f}$. This choice is made so that if randomly selected links rewire, then the flux from AA links to AB links,  $(1-f)d(1-\alpha)$, equals the flux from AB links to AA links, $\frac{f}{2}\alpha $, and likewise for balance of fluxes between BB and AB links. Therefore, if SI links occur at random
anywhere in the network, this rewiring strategy will on average keep the
community structure specified above by $d$ and $f$.


We simulate our model using Gillespie's method \cite{DanielT}  for  $N=10^{4}$ nodes  and
$K=10^{5}$ links \cite{Shaw2008a}. The initial condition is either
the final state of a previous run or a random two-community network
constructed as described above in which a fraction of the nodes have been
randomly infected.

As in \cite{Gross2006,Shaw2008a}, we derive mean field equations
for the evolution of the nodes and links. $P_X$ denotes the probability  of
nodes to be in state $X$, where $X$ is susceptible in community A or B ($S_a$
or $S_b$) or infected in A or B ($I_a$ or $I_b$). $P_{XY}$ denotes the
probability that a randomly selected link connects a node in state $X$ to a
node in state $Y$. We obtain the following equations for the node dynamics:
\begin{eqnarray}
\label{eq:mf_node}
\dot P_{Ia}&=&-rP_{Ia}+\frac{pK}{N_a}(P_{SaIa}+P_{SaIb}) \label{eq:nodeA}\\
\dot P_{Ib}&=&-rP_{Ib}+\frac{pK}{N_b}(P_{SbIa}+P_{SbIb})\label{eq:nodeB}
\end{eqnarray}
Because nodes are neither created nor destroyed and do not change their
community assignment, the equations for susceptibles in community A and B can
be found from node conservation.

The evolution of the links depends on three point terms. As in
\cite{Gross2006,Zanette2007,Shaw2008a} we use a moment
closure assumption to close the system, assuming $P_{XYZ}\approx
\frac{P_{XY}P_{YZ}}{P_{Y}}$,  where $P_{XYZ}$ is the fraction of three point
terms. After applying the moment closure, the link
equations are
\begin{eqnarray}
\label{eq:mf_link}
\dot P_{SaSa}&=&r P_{SaIa}+w \alpha (P_{SaIa}+P_{SaIb}) \nonumber\\
& &-\frac{2pK}{N_a}(\frac{P_{SaSa}P_{SaIa}}{P_{Sa}}+\frac{P_{SaSa}P_{SaIb}}{P_{Sa}})\label{eq:SaSa}\\
\dot P_{SbSb}&=&r P_{SbIb}+w \beta (P_{SbIa}+P_{SbIb})\nonumber \\
& &-\frac{2pK}{N_b}(\frac{P_{SbSb}P_{SbIa}}{P_{Sb}}+\frac{P_{SaSb}P_{SbIb}}{P_{Sb}})\label{eq:SbSb}\\
\dot P_{SaSb}&=&r P_{SbIa}+r P_{SaIb}+w(1- \alpha)(P_{SaIa}+P_{SaIb})\nonumber\\
& &+w(1-\beta)(P_{SbIa}+P_{SbIb})\nonumber\\
& &-\frac{pK}{N_a}(\frac{P_{SbSa}P_{SaIa}}{P_{Sa}}+\frac{P_{SbSa}P_{SaIb}}{P_{Sa}})\nonumber\\
& &-\frac{pK}{N_b}(\frac{P_{SaSb}P_{SbIa}}{P_{Sb}}+\frac{P_{SaSb}P_{SbIb}}{P_{Sb}})\label{eq:SaSb}\\
\nonumber\\
\dot P_{SaIa}&=&2rP_{IaIa}-(r+p+w)P_{SaIa}\nonumber\\
& &+\frac{2pK}{N_a}(\frac{P_{SaSa}P_{SaIa}}{P_{Sa}}+\frac{P_{SaSa}P_{SaIb}}{P_{Sa}})\nonumber\\
& &-\frac{pK}{N_a}(\frac{P_{SaIa}^2}{P_{Sa}}+\frac{P_{SaIa}P_{SaIb}}{P_{Sa}})\label{eq:SaIa}\\
\dot P_{SbIb}&=&2rP_{IbIb}-(r+p+w)P_{SbIb}\nonumber\\
& &+\frac{2pK}{N_b}(\frac{P_{SbSb}P_{SbIa}}{P_{Sb}}+\frac{P_{SbSb}P_{SbIb}}{P_{Sb}})\nonumber\\
& &-\frac{pK}{N_b}(\frac{P_{SbIb}P_{SbIa}}{P_{Sb}}+\frac{P_{SbIb}^2}{P_{Sb}})\label{eq:SbIb}\\
\dot P_{SaIb}&=&rP_{IaIb}-(r+p+w)P_{SaIb}\nonumber\\
& &+\frac{pK}{N_b}(\frac{P_{SbSb}P_{SbIa}}{P_{Sb}}+\frac{P_{SbSb}P_{SbIb}}{P_{Sb}})\nonumber\\
& &-\frac{pK}{N_a}(\frac{P_{SaIb}P_{SaIa}}{P_{Sa}}+\frac{P_{SaIb}^2}{P_{Sa}})\label{eq:SaIb}\\
\nonumber\\
\dot P_{SbIa}&=&rP_{IaIb}-(r+p+w)P_{SbIa}\nonumber\\
& &+\frac{pK}{N_a}(\frac{P_{SaSb}P_{SaIa}}{P_{Sa}}+\frac{P_{SaSb}P_{SaIb}}{P_{Sa}})\nonumber\\
& &-\frac{pK}{N_b}(\frac{P_{SbIa}^2}{P_{Sb}}+\frac{P_{SbIa}P_{SbIb}}{P_{Sb}})\label{eq:SbIa}\\
\dot P_{IaIa}&=&-2rP_{IaIa}+pP_{SaIa}\nonumber\\
& &+\frac{pK}{N_a}(\frac{P_{SaIa}^2}{P_{Sa}}+\frac{P_{SaIa}P_{SaIb}}{P_{Sa}})\label{eq:IaIa}\\
\dot P_{IbIb}&=&-2rP_{IbIb}+pP_{SbIb}\nonumber\\
& &+\frac{pK}{N_b}(\frac{P_{SbIb}P_{SbIa}}{P_{Sb}}+\frac{P_{SbIb}^2}{P_{Sb}})\label{eq:IbIb}\\
\dot P_{IaIb}&=&-2rP_{IaIb}+pP_{SbIa}+pP_{SaIb}\nonumber\\
& &+\frac{pK}{N_a}(\frac{P_{SaIb}P_{SaIa}}{P_{Sa}}+\frac{P_{SaIb}^2}{P_{Sa}})\nonumber\\
& &+\frac{pK}{N_b}(\frac{P_{SbIa}^2}{P_{Sb}}+\frac{P_{SbIa}P_{SbIb}}{P_{Sb}})\label{eq:IaIb}
\end{eqnarray}
These mean field equations are a special case of the general mean field in \cite{Wang2011} for appropriate choices of their rewiring and infection parameters.

Since the total number of links is fixed, we have an 11 dimensional system in
the adaptive network case by eliminating  one of the link
equations.  On the other hand, we have a 9 dimensional system in the static
network case because the numbers of AA, AB, and BB links are each fixed. These equations can be integrated with standard
numerical integration methods. Also, we tracked their steady states
using a continuation package \cite{XPPAUT}.

\section{Results}

\label{sec:results}

\subsection{Static network}

\label{subsec:static}

\begin{figure}[tbp!]
\includegraphics[width=3.8in,keepaspectratio]{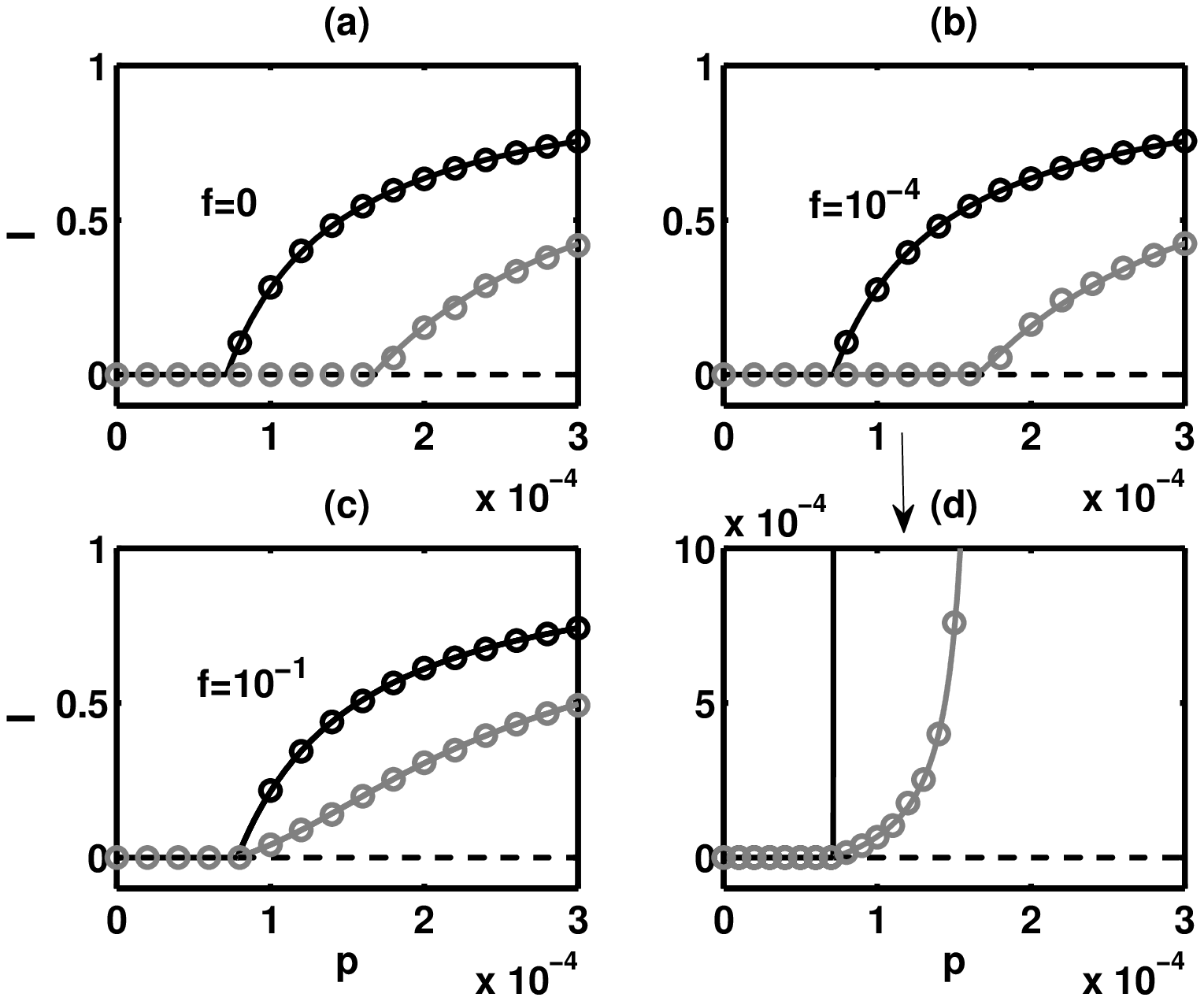}
\caption{Bifurcation diagram for infected fraction in a static network ($w=0$) as a function of infection rate $p$ for different
fractions of cross links $f$.  Black: community A; gray: community B.  Solid
curves: mean field solutions (stable branches); dashed curves:
mean field solutions (unstable branches); circles: Monte Carlo simulations.
Average degrees $\left<k_a\right>\approx 28$ and $\left<k_b\right>\approx 12$. The plots
 correspond to $N_a=5\times10^3$, $N_b=5\times10^3$, $K=10^6$, $d=0.7$ and
 $r=0.002$. In (d), in order to obtain better statistics, we averaged over 100 network realizations, each of which is
 similar to (b).
}
\label{fig:panelw0}
\end{figure}

We first consider the effect of having two communities with different average
connectivities in a static network ($w=0$).  We obtained the bifurcation structure (Figure \ref{fig:panelw0}) as follows.
We used the XPPAUT free software package \cite{XPPAUT} to locate the stable and
unstable equilibrium solutions of mean field equations. To obtain the steady
states of the full system, we generated an initial random network with
community structure in which $50\%$ of the nodes were infected. To locate the
upper branch (endemic state), the system was run to steady state for a high
infection rate $p$, and $p$ was decreased gradually using the final state of
each run as an initial state for the next run.
For each $p$, we run the system up to $5\times10^4$ time units and then averaged the
steady state over $500$ samples where there are $10^3$ events between each
sample.
To locate the lower branch (disease-free state), we generated a random network
with community structure in which $2.5\%$ of the nodes were infected. The
system was simulated for $5\times10^4$ time units, and five runs were done for
each $p$ value. If the infected fraction went to zero in any of the five runs,
the disease-free state was considered stable. As shown in Figure
\ref{fig:panelw0}, the mean field equations and the full system are in good
agreement.

In a static network without community structure, the disease free state (DFS) loses stability  at a critical
infection rate $p^*$ where the system undergoes a transcritical
bifurcation. This threshold infection rate  depends on the average degree of
the network \cite{Gross2006}. Figure \ref{fig:panelw0}a
superimposes the bifurcation diagrams of two single-community networks with
different average degrees.  The epidemic threshold for community A ($p_a^*$) and community B ($p_b^*$) are significantly different. When the
two networks are loosely connected ($f=10^{-4}$, Figure
\ref{fig:panelw0}b and blowup in Figure \ref{fig:panelw0}d), the combined
system has a single threshold infection rate, which
is approximately $p_{a}^{*}$ and much lower than $p_b^*$ in the disconnected
case. When the infection rate is  between $p_a^*$ and $p_b^*$,
the fraction of infecteds in B is close to zero (Figure \ref{fig:panelw0}d)
and stochastic reintroduction of infection from A to B is observed.
However, when the two communities are strongly connected ($f=10^{-1}$, Figure
\ref{fig:panelw0}c), they behave similarly in that both communities
have significant infection levels for the same parameter values.

\begin{figure}[tbp]
\includegraphics[width=3in,keepaspectratio]{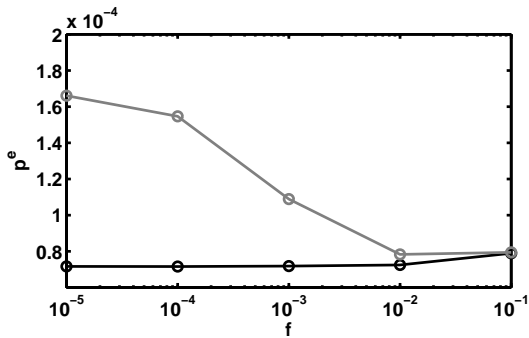}
\caption{Effective threshold infection rate from mean field vs fraction of cross links $f$. Black curve: community A; gray curve: community B.
The infection rate $p$ at which the fraction of infecteds is less than
$10^{-3}$ is considered as the effective threshold rate. $N_a=5\times10^3$, $N_b=5\times10^3$, $K=10^6$, $d=0.7$ and $r=0.002$.
}
\label{fig:thresholdw0}
\end{figure}

Although a system of two connected communities has an infection threshold at a
single bifurcation point, we wish to distinguish between the cases in Figure
\ref{fig:panelw0}b,c, where significant infection spread occurs in the low
degree community (B) if it has sufficient links to the high degree community
(A), while the infection in community B is very small if the number of cross
links is low.  To quantify this, we define effective threshold infection rates
$p^e_a$, $p^e_b$ for each community as follows. While sweeping the infection rate $p$ from higher to lower values, the first $p$ value at which the fraction of infecteds at the steady state is lower than $\epsilon=10^{-3}$ is considered as the effective threshold infection rate for that community.  Figure \ref{fig:thresholdw0} shows the effective threshold infection rates versus cross link fraction $f$.  When $f<10^{-2}$, the effective thresholds in the two communities become noticeably different.

\subsection{Adaptive network}
\label{subsec:adaptive}

We now move to systems with nonzero rewiring rates.  The bifurcation structure
for the adaptive network case was determined as for static networks except for the following modification. In our model, network
adaptation does not occur in the absence of infection.  This means that the
DFS of Equations (\ref{eq:nodeA}-\ref{eq:IaIb}) is not isolated, because any
disease free combination of AA, AB, and BB links is a steady state. Because
of the non-isolated fixed points, the stability of the disease free branch
could not be determined using continuation packages. Instead, we calculated
numerically the eigenvalues of the Jacobian evaluated at the DFS for the
initial network geometry which is described by $f$ and $d$.

\begin{figure}[tbp]
\includegraphics[width=3.7in,keepaspectratio]{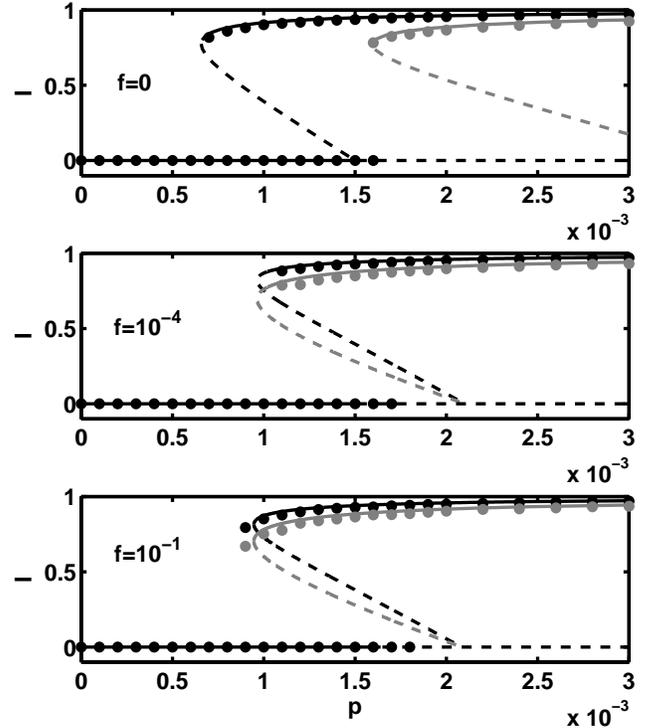}
\caption{Bifurcation diagram for infected fraction in an adaptive network  as a function of infection rate $p$ for different
fractions of cross links $f$. Black: community A; gray: community B.  Solid curves: mean field solutions (stable branches); dashed curves: mean field solutions (unstable branches); circles: Monte Carlo simulations.
Average degrees $\langle k_a \rangle \approx 28$ and $\langle k_b \rangle\approx 12$. The plots
correspond to $N_a=5\times10^3$, $N_b=5\times10^3$, $K=10^6$, $d=0.7$,
$w=0.04$ and $r=0.002$.
The simulations are done similarly as in Figure \ref{fig:panelw0}, but here we
run the system for  $5\times 10^6$ time units in order to approach the endemic branch.}
\label{fig:panelw04}
\end{figure}

In the absence of community structure ($f=0$), the DFS loses stability at a critical infection rate $p^*$,
where the unstable endemic branch and the stable disease free branch
intersect at a transcritical bifurcation (Figure \ref{fig:panelw04}a). The epidemic
threshold $p^*$ is inversely proportional to the average degree of the
network and can be found analytically from the Jacobian of the mean-field
equations for a single network (see Appendix).

For a network having two loosely connected heterogenous communities
($f=10^{-4},d=0.7$), $p^*$ is very close to that of a single network having the same average degree as community A
(Figure \ref{fig:panelw04}a,b). This is expected because when the infection in
community A starts to spread, a small value of $f$ will not be enough to stop
infection spreading in community A. In contrast, for $f=10^{-1}$, $p^*$ is
larger than that of a single network having the same average degree as community A.
In this case, the connection is stronger and the infection starting
to spread in community A can be suppressed by the connection to a community
where no infection is observed. As we increase $f$, the
critical value of $p$ approaches that of a single network with average
degree $\langle k \rangle=20$, which is the average degree in the entire system.


In an adaptive network without community structure, bistability can occur for
a range of rewiring rates \cite{Gross2006}. We focus on the rewiring rate
$w=0.04$, for which the endemic branch loses stability at a critical infection
rate where the system undergoes a saddle-node (SN) bifurcation.
The location of this SN bifurcation depends on the average degree of the
network. Figure \ref{fig:panelw04}a
superimposes the bifurcation diagrams of two single-community networks with
different average degrees.

In our model, for a network having two strongly connected
communities ($f=10^{-1}$), the endemic state loses stability at
a SN bifurcation point. However, for the loosely connected case ($f=10^{-4}$), the endemic
branch loses stability at a Hopf bifurcation (HB) point.
As with the epidemic threshold (transcritical bifurcation) in static networks,
the location of the SN bifurcation and the HB in the adaptive network is
governed primarily by the high degree community. However, in contrast with the
static case, small cross link fraction $f$ is not associated with low steady state
infection levels in the low degree community.  As will be seen later in this section,
the state with high infection in one community and low in the other community is not a
steady state due to the network adaptation.  Instead, infection levels are
similar in both communities even if weakly connected.
Both communities continue to exhibit high infection levels as the number of cross links is further increased (Figure \ref{fig:panelw04}c).

By looking at the time series of the mean field equations, we observed  a stable periodic
solution for a very small range of $p$ values near the HB point. However, we do not
see any periodic behavior in the full system due to the narrow range of $p$ values.

As the heterogeneity in the network increases by increasing $d$, the Hopf
bifurcation point also increases.
Furthermore, the epidemic threshold decreases because of a much higher average
degree in community A. This narrows the region where both
disease free and endemic branches are stable. In
particular, for $d=0.9$, there is no $p$ value where both the endemic
steady state and DFS are stable. For $d=0.9$ We observed periodic solutions with a
very long period for $p$ values smaller than the HB point in both mean field
and the full system.

\begin{figure}[tbp]
\includegraphics[width=3.7in,keepaspectratio]{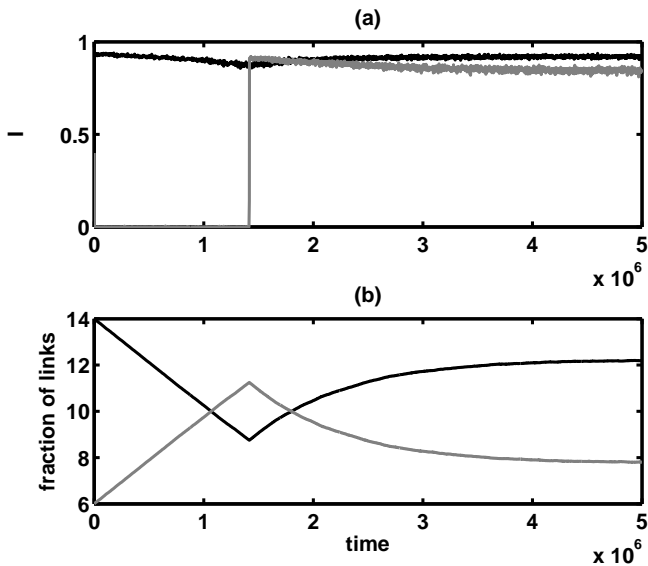}
\caption{Time series from Monte Carlo simulation of full system. (a) Infected
  node fractions. Black curve: fraction of infecteds in community A; gray
  curve: fraction of  infecteds in community B. (b) Fraction of links. Black
  curve: fraction of AA type links; gray curve: fraction of  BB type
  links. Initial condition is as described in the text. $N_a=5\times10^3$,
  $N_b=5\times10^3$, $K=10^6$, $f=10^{-4}$, $d=0.7$, $w=0.04$, $r=0.002$ and
  $p=0.0014$.}
\label{fig:timeSd07}
\end{figure}

To motivate the absence of the state seen in static networks with high
infection in one community and low infection in the other, we consider a long
time series starting initially with a network where $f=10^{-4}$ and $d=0.7$
generated as described above.
As seen in Figure \ref{fig:timeSd07}, the fraction of AA and BB type links
change with time.  The adaptation rules have been chosen so that if SI links
are distributed uniformly throughout the network, the community structure will
be preserved.  However, with high initial infection levels in A and low in B,
there are more SI links among the AA and AB links and fewer among the BB
links.  This leads to a net flux of link types from AA to AB to BB.  (The
fraction of AB links (not shown) remains relatively constant at low levels
throughout.)  Eventually the average degree in the B community exceeds that in
the A community and there is an incursion of infection from A to B.  The flux
of link types is then reversed, and the steady state network structure is
similar (but not identical) to that expected from the community structure
parameters $d,f$.  The infection persists at high levels in both communities
at steady state.

\begin{figure}[tbp]
\includegraphics[width=3.7in,keepaspectratio]{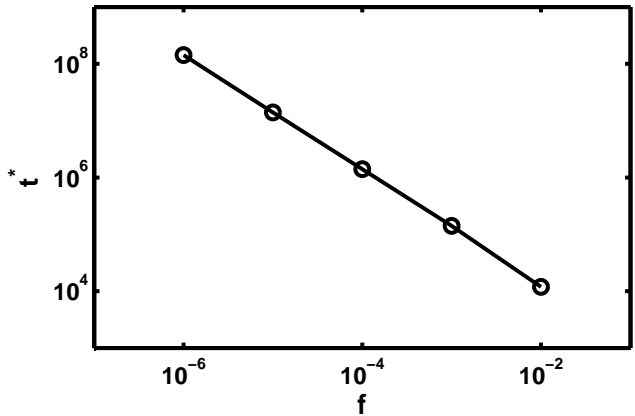}
\caption{The critical time for infection incursion in community B vs
 fraction of cross links $f$.
The critical time values are averaged over 5 runs except for $f=10^{-6}$. In each
run, the time when the fraction of infecteds in community B exceeded $0.1\%$ is considered
the critical time.}
\label{fig:tvsf}
\end{figure}

We can estimate the time until infection incursion in the B community as
follows.
From Equations (\ref{eq:SaSa},\ref{eq:SaIa},\ref{eq:IaIa}), the fraction of AA links $P_{aa}=P_{SaSa}+P_{SaIa}+P_{IaIa}$ evolves according to
\begin{equation}
\dot P_{aa}=w\left[\alpha P_{SaIb}-(1-\alpha)P_{SaIa}\right].
\end{equation}
Since we are interested in the
critical time when infection starts to spread in community B, we can assume
$P_{Ib}\approx 0$ and hence $P_{SaIb} \approx 0$.  Thus
\begin{eqnarray}
\label{eq:mf_AA}
\dot P_{aa}& \approx &-w(1-\alpha)P_{SaIa}\nonumber \\
&\approx&-w(1-\alpha)\gamma P_{aa}
\end{eqnarray}
where $\gamma=P_{SaIa}/P_{aa}$.  Since $f$ is close to zero, we can use the single mean field
equations to approximate $\gamma$, the fraction of SI links in community A (see Appendix A). We can then solve Equation (\ref{eq:mf_AA}) for the critical time $t_c$ for infection incursion if we
know  $P_{aa}$ at that time. Since we can predict the critical average degree for community B in order
for the disease to spread, we can also find $P_{aa}$ at that point. However, for the full system, the infection in community B starts to spread
much earlier than the time found by using mean field equations because of the stochastic nature of our model. Even so, we can solve the equation for $t$
as follows:
\begin{eqnarray}
t&=& \ln\left[\frac{P_{aa}(0)}{P_{aa}(t)}\right]\left[w(1-\alpha)\gamma\right]^{-1}\\
\end{eqnarray}
from which it can be shown that the critical time $t_c$ is proportional to
$\frac{1}{f}$. In Figure \ref{fig:tvsf},
we can see that the relationship $t_c \propto \frac{1}{f}$ holds for the full system.

\begin{figure}[tbp]
\includegraphics[width=3.5in,keepaspectratio]{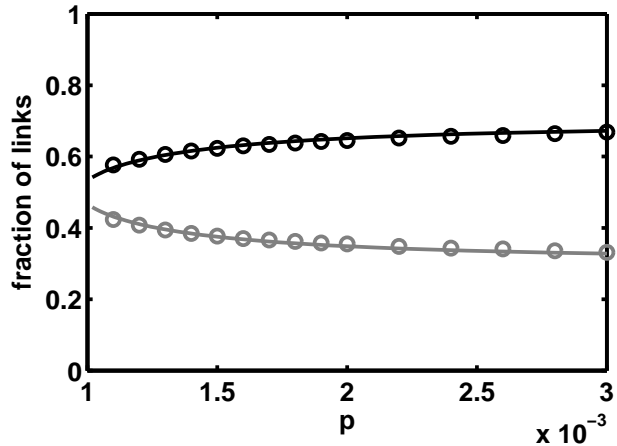}
\caption{Fraction of links vs infection rate $p$.
Curves: mean field steady states;
points: Monte Carlo steady state averages.
Black:  fraction of AA type
links; gray:  fraction of BB type links.
(The fraction of AB links is less than $1\%$.) $N_a=5\times10^3$, $N_b=5\times10^3$,
$K=10^6$, $f=10^{-4}$, $d=0.7$, $w=0.04$, $r=0.002$.
The data is  from the simulations done in  Figure \ref{fig:panelw04}.
}
\label{fig:linksw04}
\end{figure}

When the system reaches steady state (late time in Figure \ref{fig:timeSd07}),
the network geometry does not return exactly to that expected from the
community structure parameters $f,d$ because the SI links are not uniformly
distributed.  This effect is most pronounced when the infection rate is below
the critical infection rate for the low connectivity network, because then the
SI link distribution is the most nonuniform.  Figure \ref{fig:linksw04} shows
the steady state community structure versus the infection rate.  Deviations
from the community structure specified by $f,d$ increase as the infection rate
approaches the Hopf bifurcation point at $p\approx 0.0019$.  Thus the steady state average degree observed in the two communities in the presence of an epidemic can be different than that expected in the absence of an epidemic.  The adaptation has a homogenizing effect, bringing the degrees in the communities closer to each other.

\begin{figure}[tbp]
\includegraphics[width=3.7in,keepaspectratio]{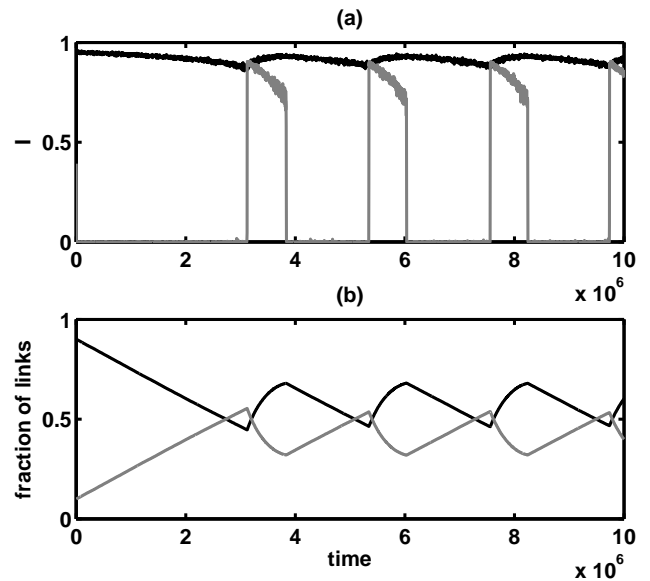}
\caption{Time series from Monte Carlo simulation of full system. (a) Infected
  node fractions. Black curve: fraction of infecteds in community A; gray
  curve: fraction of  infecteds in community B. (b) Fraction of links. Black
  curve: fraction of AA type links; gray curve: fraction of  BB type
  links. Initial condition is as described in the text. $N_a=5\times10^3$,
  $N_b=5\times10^3$, $K=10^6$, $f=10^{-4}$, $d=0.9$, $w=0.04$, $r=0.002$ and
  $p=0.0014$.}
\label{fig:timeSd09}
\end{figure}

For $d=0.9$, the behavior of the system is very interesting. In
Fig.~\ref{fig:timeSd09}, we started with similar initial conditions as for
$d=0.7$ in Fig.~\ref{fig:timeSd07}. The system behaved similarly for a long time, but once the infection
level reached a high value in community B, it could not stay in that state,
because a higher $d$ means a higher out flux rate from BB type links. The
average degree in B started to decrease, causing infection to die out again in
B. A periodic solution with a long period is observed for a range of $p$
values.

\section{Conclusions}

\label{sec:conclusions}

We have studied epidemic spread in a network of two communities with different average degrees.  Cases with and without disease avoidance rewiring were considered.  Rewiring rules were chosen so that the community structure would be preserved if links between susceptibles and infectives occurred uniformly throughout the network.  The steady state bifurcation structure was obtained for static and adaptive cases.  A mean field theory based on a moment closure approximation accurately predicted the steady state infection levels and network structure observed in stochastic simulations of the full model.

In the static network case, weakly connected communities displayed significantly different infection levels.  Low infection levels could persist in a subthreshold community weakly connected to a high degree, high infection community.  Increasing the number of connections between communities led to more similar behavior of the two communities.  In contrast, communities in adaptive networks displayed similar infection levels even if weakly connected.  Steady states with high infection in one community and low in the other did not exist for the adaptive network case.

The absence of steady states with significantly different infection levels was explained by considering network adaptation in the presence of nonuniformly distributed SI links.  If one community has few infectives, there is a net flux of links into that community until its degree is high enough to support the infection.  We estimated the time until this infection incursion based on mean field arguments and found that the time increases as the number of cross links between communities decreases.

We also observed changes in the steady state network geometry due to adaptation in the presence of infection.  These changes were most significant near bifurcation points.  The adaptation tended to bring the average degrees of the communities closer to each other.  Thus the adaptation promotes greater similarity between communities in both network structure and infection levels.

The model presented in this paper is the first to include community structure in epidemic spread on an adaptive network.  Future work is needed to extend the model to more realistic scenarios.  For example, the number of communities could be increased beyond two.  We have observed that the convergence time to steady state can be very long for weakly coupled communities, so it is possible that an epidemic would not reach steady state during physically realistic time scales.  Thus the transient behavior should also be studied in more detail.  Identifying when communities become at risk for incursion of infection could be valuable in knowing when epidemic control measures are needed.  Another area for future extension is to change the rules for adaptation, such
as cutting or temporarily deactivating links rather than rewiring them.

This work was supported by the Army Research Office, Air Force Office of Scientific Research, and by Award Number R01GM090204 from the National Institute Of General Medical
Sciences.   The content is solely the responsibility of the authors and does not necessarily
represent the official views of the National Institute of General Medical Sciences or
the National Institutes of Health.

%

\appendix

\section{Analytical solution for a single community}

For a single community ($f=0$), the mean field equations are:
\begin{eqnarray*}
\label{eq:app}
\dot P_{I}&=&-rP_{I}+\frac{pK}{N}P_{SI}\\
\dot P_{SI}&=&2r(1-P_{SI}-P_{SS})-(r+p+w)P_{SI}\\
& &+\frac{2pK}{N}\frac{P_{SS}P_{SI}}{P_{S}}-\frac{pK}{N}\frac{P_{SI}^2}{P_{S}}\\
\dot P_{SS}&=&r P_{SI}+wP_{SI}\\
& &-\frac{2pK}{N}\frac{P_{SS}P_{SI}}{P_{S}}
\end{eqnarray*}
At steady state, we obtain
\begin{eqnarray}
\label{eq:app}
& &-rP_{I}+\frac{pK}{N}P_{SI}=0 \label{eq:single_I}\\
& &2r(1-P_{SI}-P_{SS})-(r+p+w)P_{SI}\nonumber\\
& &+\frac{2pK}{N}\frac{P_{SS}P_{SI}}{P_{S}}-\frac{pK}{N}\frac{P_{SI}^2}{P_{S}}=0\label{eq:single_SI}\\
& &rP_{SI}+wP_{SI}-\frac{2pK}{N}\frac{P_{SS}P_{SI}}{P_{S}}=0 \label{eq:single_SS}
\end{eqnarray}

To find the endemic steady state, we first solve [\ref{eq:single_I}] for $P_{SI}$, and substitute into
[\ref{eq:single_SI}]. Then we solve [\ref{eq:single_SI}] for $P_{SS}$ in terms
of $P_I$. After substituting that into [\ref{eq:single_SS}], we obtain
a quadratic equation in $P_I$, $AP_I^2+BP_I+C=0$, where
\begin{eqnarray*}
A&=&p-w \\
B&=&2w-p-2p\frac{K}{N} \\
C&=&2p\frac{K}{N}-w-r\\
\end{eqnarray*}
The quadratic can be solved analytically for $P_I$, and then $P_{SI}$ and the other link variables can be computed.

\bibliographystyle{apsrev}

\end{document}